\documentclass[amssymb, preprintnumbers, showpacs, showkeys, aps, prl, superscriptaddress, twocolumn, longbibliography]{revtex4-1}

\usepackage[hidelinks, hypertexnames=false]{hyperref}

\usepackage[all]{hypcap}
\usepackage{xcolor}
\usepackage{slashed}
\usepackage{graphicx}
\usepackage{amsmath}
\usepackage{latexsym}
\usepackage{epstopdf}
\usepackage{enumitem}
\usepackage{amssymb}
\usepackage{multirow}
\usepackage{mathtools}
\usepackage{bbm}
\usepackage{bm}
\usepackage{amsfonts}
\usepackage{amssymb}
\usepackage{calligra}
\usepackage{calrsfs}
\usepackage{makecell}
\usepackage[normalem]{ulem}
\usepackage[USenglish,american]{babel}
\usepackage{physics}
\usepackage{braket}
\usepackage{cancel} 

\newcommand{\be}{\begin{equation}}
\newcommand{\ee}{\end{equation}}
\newcommand{\ba}{\begin{eqnarray}}
\newcommand{\ea}{\end{eqnarray}}
\newcommand{\bs}{\begin{split}}
\newcommand{\es}{\end{split}}

\expandafter\ifx\csname package@font\endcsname\relax\else
 \expandafter\expandafter
 \expandafter\usepackage
 \expandafter\expandafter
 \expandafter{\csname package@font\endcsname}%
\fi

\begin{document}

\title{Exploring Quantum Phases of Dipolar Gases through Quasicrystalline Confinement
}

\author{Vinicius Zampronio}%
\email{v.zamproniopedroso@unifi.it}
\affiliation{Departamento  de  F\'isica  Te\'orica  e  Experimental,  Universidade  Federal  do  Rio  Grande  do  Norte, 59075-000 Natal, Brazil}
\affiliation{Dipartimento di Fisica e Astronomia, Universit\`a di Firenze, I-50019, Sesto Fiorentino (FI), Italy}

\author{Alejandro Mendoza-Coto}%
\email{alejandro.mendoza@ufsc.br}
\affiliation{Departamento de F\'\i sica, Universidade Federal de Santa Catarina, 88040-900 Florian\'opolis, Brazil}%
\affiliation{Max Planck Institute for the Physics of Complex Systems, Nothnitzerstr. 38, 01187 Dresden, Germany}

\author{Tommaso Macr\`i}%
\email{tommasomacri@gmail.com}
\affiliation{ITAMP, Harvard-Smithsonian Center for Astrophysics, Cambridge, Massachusetts 02138, USA}
\affiliation{Departamento  de  F\'isica  Te\'orica  e  Experimental,  Universidade  Federal  do  Rio  Grande  do  Norte, 59075-000 Natal, Brazil}

\author{Fabio Cinti}
\email{fabio.cinti@unifi.it}
\affiliation{Dipartimento di Fisica e Astronomia, Universit\`a di Firenze, I-50019, Sesto Fiorentino (FI), Italy}
\affiliation{INFN, Sezione di Firenze, I-50019, Sesto Fiorentino (FI), Italy}
\affiliation{Department of Physics, University of Johannesburg, P.O. Box 524, Auckland Park 2006, South Africa}

\begin{abstract}
The effects of frustration on extended supersolid states is a largely unexplored subject in the realm of cold-atom systems.
In this work, we explore the impact of quasicrystalline lattices on the supersolid phases of dipolar bosons. Our findings reveal that weak quasicrystalline lattices can induce a variety of modulated phases, merging the inherent solid pattern with a quasiperiodic decoration induced by the external potential. As the lattice becomes stronger, we observe a super quasicrystal phase and a Bose glass phase. 
Our results, supported by a detailed discussion on experimental feasibility using dysprosium atoms and quasicrystalline optical lattice potentials, open a new avenue in the exploration of long-range interacting quantum systems in aperiodic environments. We provide a solid foundation for future experimental investigations, potentially confirming our theoretical predictions and contributing profoundly to the field of quantum gases in complex external potentials.
\end{abstract}

\maketitle


\textit{Introduction} -- The exploration of exotic quantum phases displaying simultaneously different kinds of orders, or \textit{quasi}-orders in systems with long-range interactions, has emerged as a central goal in condensed-matter physics~\cite{Defenu2023}. Among these, electronic liquid crystals~\cite{Kivelson1998,Fernandes2014}, Bose glasses (BGs)~\cite{Yu2012a,Yu2023}, supersolids~\cite{Boninsegni2012,Macri2013,Cinti2014,Recati2023}, and quantum systems with quasiperiodic order~\cite{Gopalakrishnan2013,Ciardi2023_1,Pupillo2020,Coto2022,grossklags2023} stand out as intriguing manifestations combining translational, rotational, and global gauge symmetry breaking. Such systems exhibit exotic collective behaviors arising from intricate interplays between quantum effects, topology, and interactions, holding promise for applications ranging from quantum simulators to advanced materials~\cite{Jaksch2005,Daley2011,Gonzalez2018,Schafer2020,Recati2023,1367-2630-16-3-033038,Wachtler2016,Wachtler2016_1}.
In this context, the study of dipolar bosons has garnered increasing interest, driven by advancements in both theoretical~\cite{Lahaye2009,Lu2015,Baillie2018,doi:10.7566/JPSJ.85.053001,PhysRevLett.105.135301,nho2005,PhysRevA.96.013627,Kora2019,PhysRevLett.119.215302,zhang2019,Roccuzzo2019,Halperin2023,Ciardi2024} and experimental~\cite{Li2017,Leonard2017,Tanzi2019,Tanzi2019-2,Bottcher2019,Chomaz2019,Guo2019,Natale2019,Norcia2021,Chomaz2022,Biagioni2024} research.

Recent investigations in these systems revealed a rich phase diagram with three distinct supersolid phases—stripes, triangular, and honeycomb—eventually converging at a critical point~\cite{zhang2019,Ripley2023,Zhang2023}.
In the conventional scenario, the crystallization transition is expected to be first-order, resulting in a sudden increase in the modulation amplitude upon dynamically crossing the crystallization line. This impacts the phase coherence of the system and leads to the generation of high-energy excitations~\cite{Kadau2016}, hindering the stability of the supersolid phase.

\begin{figure}[t!]
    \centering
    \includegraphics[width=\linewidth]{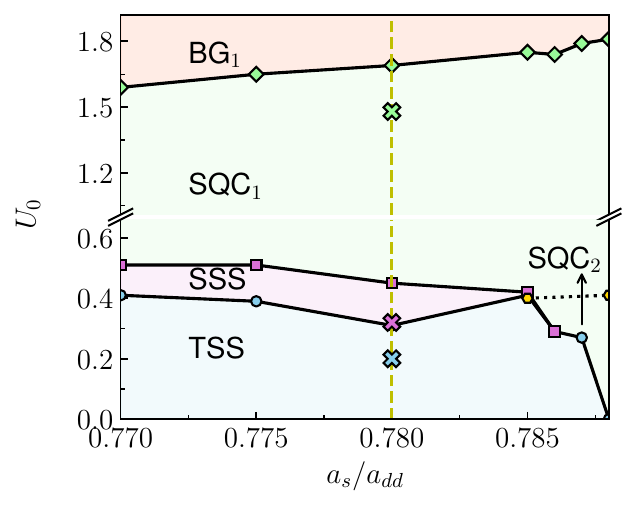}
    \caption{{\textbf{Phase diagram of dipolar bosons in a quasicrystalline lattice.}} 
    We numerically solve Eq.~\eqref{eq1} in imaginary time to demonstrate the emergence of several phases induced by the simultaneous effect of a quasicrystalline potential with intensity $U_0$ and contact (with strength $a_s$) and dipolar ($a_{dd}$) interactions.
    We observe the triangular-supersolid (TSS), stripe-supersolid (SSS), super-quasicrystal (SQC$_{i=1,2}$), and Bose glass (BG$_1$) phases, as described in the text. 
    The average scaled density of the effective two-dimensional system is $\bar{\rho}_{\perp}=120$, the wave vector of the QCL is $q$ = 1.8,
    and the frequency of the harmonic trap along the polarization axis is $\omega_z = 0.08$.
    Solid lines delimit the phase boundaries. The dotted line guides the eye for the crossover between different SQC structures.
    The vertical dashed line corresponds to the analysis in Fig.\ref{fig3}. The crosses are representative points of the density patterns in Fig.\ref{fig2}.}
    \label{fig1}
\end{figure}
\begin{figure}[t!]
    \centering
   \includegraphics[width=\linewidth]{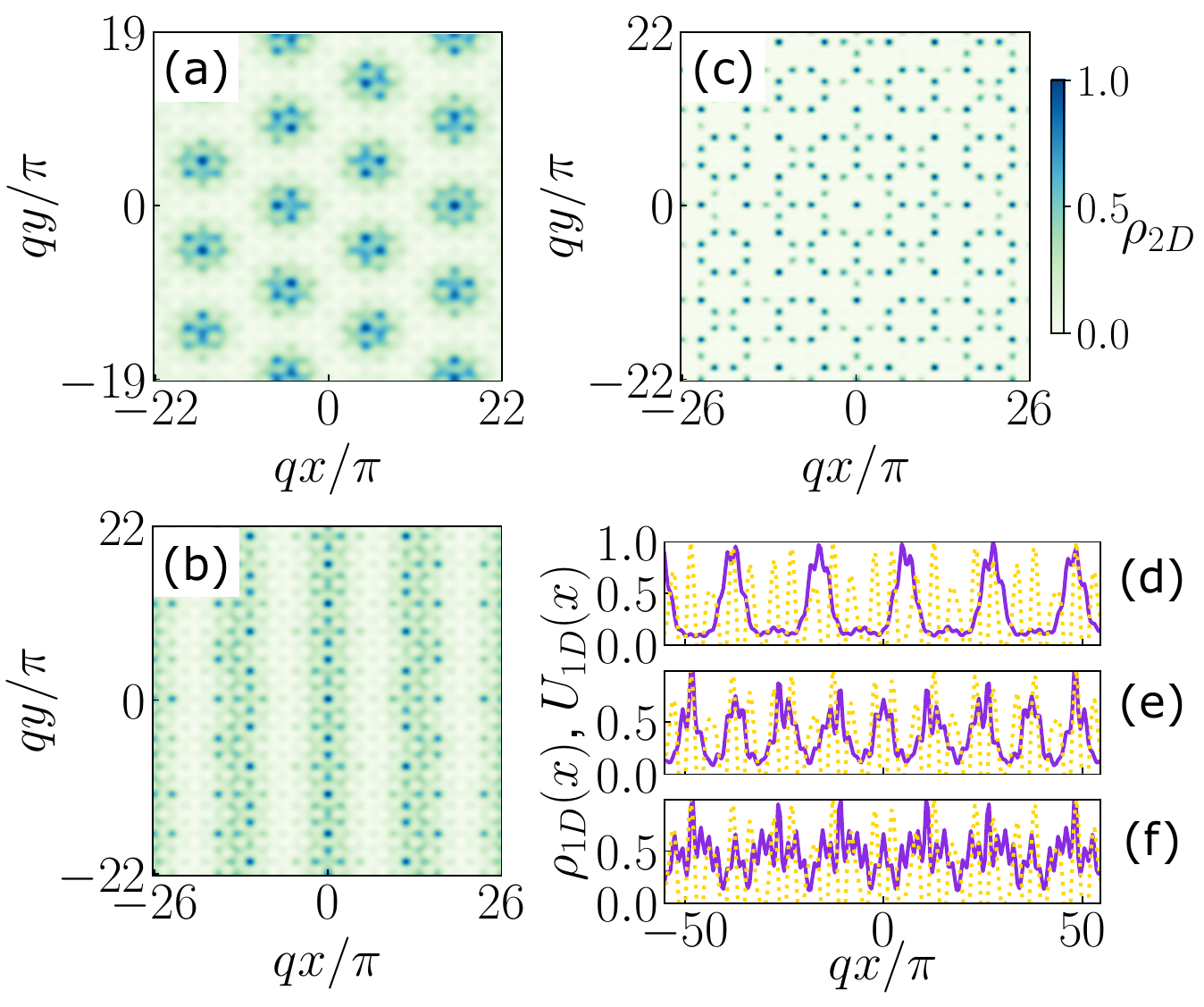}
    \caption{ \textbf{Normalized density patterns.}(a)-(c) Two-dimensional density $\rho_{2D}=\rho_{\perp}(x,y)/\rho^{\max}_{\perp}$ obtained via the numerical solution of Eq.(\ref{eq1}).
    (d)-(f) $\rho_{1D}(x)=\rho_{\perp}(x,0)/\rho_{\perp}^{\max}$ (solid curves) and QCL $U_{1D}(x)=U_q(x,0)/U_q^{\max}$ (dotted curves), at the $y = 0$ line.
    The system parameters are the same as in Fig.~\ref{fig1} with $a_s/a_{dd}=0.78$.
    In (a, d), the system is in the TSS state with $U_0 = 0.20$. 
    In (b, e), the SSS state is represented for $U_0 = 0.32$. 
    In (c, f), the SQC$_1$ state is shown for $U_0 = 1.44$. 
    The BG$_1$ state is visually indistinguishable from the SQC$_1$.
    See density patterns (a)-(c) amplified in the SM~\cite{sm}.}
    \label{fig2}
\end{figure}

Exploring the introduction of disorder into these systems could further open new avenues for understanding pioneering complex phases.
In non-interacting 2D models, disorder generates non-trivial effects such as Anderson localization (AL)~\cite{Anderson1958,Roati2008,Modugno2010}. Moreover, the presence of contact interactions in disordered Bose systems frustrates AL and leads to the emergence of BG physics~\cite{Fisher1989,Yu2023} and many-body localization~\cite{RevModPhys.91.021001}.
Quasicrystal lattices (QCLs) present intriguing platforms for investigating disorder-induced phenomena, particularly due to the observed phase transition from extended to exponentially localized states with increasing QCL intensity~\cite{Roati2008,Szabo2020,Yu2023}.
In QCL environments, quantum gases with contact interactions exhibit not only the anticipated disordered system effects~\cite{Yao2019, Viebahn2019, Sbroscia2020, Yao2020, Gautier2021, Ciardi2022, Zhu2023, Zhu2023_1}, but also the emergence of quasicrystalline superfluids~\cite{Ciardi2023_1}.
In the case of long-range interacting dipolar systems, an intriguing possibility is the study of frustrated phases resulting from the competition between the natural tendency of the system to form periodic supersolids and the presence of a QCL. 

In this letter, we bridge the physics of long-range dipolar bosons with the effects of QCL confinement. 
By numerically solving extensive Gross-Pitaevskii equations (GPEs), we explore the phase diagram of the model guided by a spectral variational study. We observe the emergence of global-ordered phases, whose formation mechanism is triggered by the local disorder generated by the QCL.
Importantly, our analysis guides experimental implementations within the scope of current technological capabilities. This approach offers the potential to probe and explore these intriguing phases of matter, significantly enhancing our understanding of quantum systems in complex potentials.

\textit{Model and Methods} -- 
We consider a gas of $N$ interacting bosons of mass $m$ and dipolar length $a_{dd}$ at $T=0$. The atoms are harmonically trapped along the polarization axis and subjected to a QCL~\cite{Viebahn2019, Sbroscia2020}
in the plane perpendicular to the polarization axis. 
The condensate wave function, $\psi(\textbf{r})$, is normalized to unity; hence the local atomic density is given by $\rho(\textbf{r})=N\vert\psi(\textbf{r})\vert^2$. 
Choosing the units of length as $\ell=12\pi a_{dd}$ and time as  
$t_0=m\ell^2/\hbar$,
we can express the GPE~\cite{dalfovo1999,leggett2001} as

\begin{align}\label{eq1}
    i\dfrac{d\psi(\mathbf{r})}{dt} =\Bigg[ &-\dfrac{1}{2}\nabla^2 + U(z) +U_{q}(\mathbf{r})\nonumber \\ 
    +& \gamma N^{3/2}|\psi(\textbf{r})|^{3} +\dfrac{a_{s}}{3a_{dd}}N|\psi(\textbf{r})|^{2} \nonumber\\
    +& \dfrac{N}{4\pi}\int d\mathbf{r'}V(\mathbf{r}-\mathbf{r'})|\psi(\mathbf{r'})|^{2} \Bigg]\psi(\mathbf{r})\,.
\end{align}
The first term on the right-hand side of Eq.~\eqref{eq1} pertains to the kinetic energy, while the second term corresponds to the axial trapping potential $U(z)=\omega_z^2 z^2/2$, $\omega_z$ being the frequency of the axial trap. The in-plane octagonal QCL is described by $U_q(\textbf{r})=\sum_{i=0}^3U_0(1-\cos(\textbf{q}_i\cdot\textbf{r}))/2$,
with characteristic wave vectors $\textbf{q}_i=q(\cos(i\pi/4),\sin(i\pi/4),0)$ \footnote{It is important to notice that due to the presence of the non-local dipolar interaction, different values of the characteristic wave vector $q$ can produce different many-body ground-state phases.}.
The fourth contribution to Eq.~\eqref{eq1} corresponds to the Lee-Huang-Yang (LHY) correction to the mean-field approximation \cite{Lee1957, Lee1957_1,PhysRevLett.115.155302}, 
where $\gamma=\frac{4}{3\pi^2}(\frac{a_s}{3a_{dd}})^{5/2}[1+\frac{3}{2}(\frac{a_s}{a_{dd}})^2]$~\cite{SCHTZHOLD2006, Bisset2016}. Finally, the last two terms of Eq.~\eqref{eq1} account for the scaled contact-interaction, with scattering length $a_s$, and 
the dipolar interaction, $V(\textbf{r})=(1-\frac{3z^2}{r^2})\frac{1}{r^3}+\frac{8\pi}{3}\delta(r)$.
The ground state $\psi_0$ of the system is obtained by evolving an initial ansatz $\psi(\mathbf{r},t=0)$ in imaginary time. Details about $\psi(\mathbf{r},t=0)$ are described in the Supplemental Material (SM)~\cite{sm}.

\begin{figure}[t!]
    \centering
    \includegraphics[width=0.8\linewidth]{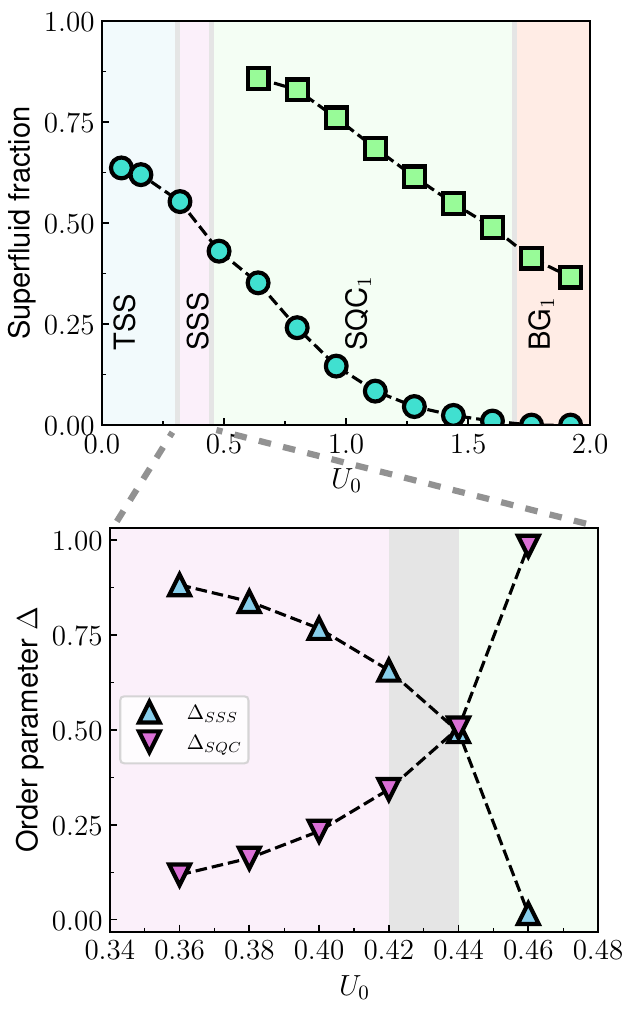}
    \caption{{\textbf{Characterization of the system ground-state.}} 
    We display the global superfluid fraction (circles), the local superfluid fraction (squares), and the SSS and SQC order parameters, $\Delta_{SSS}$ and $\Delta_{SQC}$, respectively.
    The gray regions delimit the phase transitions.
    The system parameters are the same as in Fig.~\ref{fig2}.
    }
    \label{fig3}
\end{figure}

We have also implemented a spectral variational method (SVM), see~\cite{sm} and~\cite{zhang2019, Coto2022,Grossklags20232023}. 
In summary, after introducing the average $2$D  density of particles $\rho_{\perp}(\mathbf{r}_{\perp})= N\int dz |\psi_0|^2(\mathbf{r})$, an effective energy functional in terms of $\rho_{\perp}(\mathbf{r}_{\perp}) = N|\psi_{\perp}|^2(\mathbf{r}_{\perp})$ is constructed and used to study the extended phases of the system~\cite{sm}.
The SVM shows that the optimal orientation of the periodic structure of the pattern
with respect to one of the main directions of the QCL, is $\pi/8$. 
Such a configuration is challenging to implement in the GPE solution, considering the constraints imposed by the periodic boundary conditions. However, our analytical study shows that considering two main directions of the periodic pattern and the QCL, respectively aligned, have a negligible energy cost, making numerical results with aligned patterns an excellent approximation of the actual ground state of the system~\cite{sm}.

\textit{Results.} --  
We numerically solve the GPE considering $\omega_z = 0.08$, average in-plane density $\bar{\rho}_{\perp}=120$, and $0.77 \leq a_s/a_{dd} \lesssim 0.79$.
The characteristic wave vectors of the QCL have moduli $q=1.8$, which is about five times the characteristic wave vector $k_0$ of the triangular supersolid (TSS) found when $U_0=0$.
As shown in the phase diagram in Fig.~\ref{fig1}, which is constructed by tuning the QCL intensity $U_0$ and $a_s/a_{dd}$, we identify five distinct phases in the regime of parameters considered: a TSS; a stripe supersolid (SSS); superquasicrystal phases (SQC$_{i=1,2}$), where the quasicrystalline density pattern coexists with a finite superfluid fraction, and a Bose glass (BG$_{i=1}$). Interestingly, we observe a crossover between a quasicrystalline pattern matching the external lattice ($i=2$) and a state in which it does not ($i=1$). As an effect of the long-range repulsion, many lattice sites can be depleted, resulting in a density pattern skipping a significant fraction of the QCL minima, see SM. On the other hand, when $a_s/a_{dd}$ is large, and $U_0$ is weak, the delocalization of the pattern is enhanced, and the system undergoes a crossover to a SQC state matching the QCL.
The dotted line in Fig.~\ref{fig1} is a guide to the eye for the crossover region separating the SQC$_1$ and SQC$_2$ states.
It is obtained from an analysis of the local maxima structure in $\tilde{\psi}_{\perp}(\mathbf{k}_{\perp}) \propto \int d\mathbf{r}_{\perp} \psi_{\perp}(\mathbf{r}_{\perp})e^{i\mathbf{k}_{\perp}\cdot\mathbf{r}_{\perp}}$. For this observable, two sets of local maxima can be identified, those located at $|\mathbf{k}_{\perp}| = q$, corresponding to Fourier modes matching the QCL and those located at $|\mathbf{k}_{\perp}| \neq q$, responsible for producing a quasicrystalline pattern not matching the QCL.
The position of the dotted line defines the moment when the sum of the main local maxima in $\tilde{\psi}_{\perp}(\mathbf{k}_{\perp})$ with $k_\perp\neq q$ and $k_\perp= q$ equals, i.e. $\sum_{\textbf{k}_\perp=\textbf{k}_m} \tilde{\psi}_{\perp}(\textbf{k}_\perp)\delta(k_\perp,q)=\sum_{\textbf{k}_\perp=\textbf{k}_m} \tilde{\psi}_{\perp}(\textbf{k}_\perp)[1-\delta(k_\perp,q)] $, where $\{\textbf{k}_m\}$ stand for the set of momenta of the local maxima in $\tilde{\psi}_{\perp}(\mathbf{k}_{\perp})$ and $\delta(a,b)$ corresponds to the Kronecker delta function. As can be observed in Fig~\ref{fig1}, for large enough $U_0$, the depletion effects on lattice minima produced by the dipolar interaction are always strong enough, avoiding us seeing a transition from SQC$_2$ to BG$_2$. Nevertheless, we expect this to be the case for large $a_s/a_{dd}$, as reported in Ref.~\cite{Gautier2021} for the case of contact-interacting bosons.

Remarkably, we observe the TSS-to-SSS phase transition although the SSS is never the ground state of the system at $U_0=0$ for the set of parameters considered~\cite{Ripley2023,Zhang2023}. Indeed, by increasing $U_0$ we promote the localization of particles at the minima of the QCL, inducing an increment of the size of the clusters and of the dipolar repulsion between clusters, favoring the transition to the SSS phase.
The TSS-to-SSS transition was already observed in the absence of the QCL~\cite{Ripley2023}. In this case, the enlargement of the cluster sizes is achieved by increasing the system density.

The obtained phase diagram can be understood with the help of Fig.~\ref{fig2}, where we depict the observed ground-state configurations. For weak QCLs, the ground state corresponds to the TSS state shown in Fig.~\ref{fig2}(a). As $U_0$ is increased, we see a transition to the SSS state, Fig.~\ref{fig2}(b). In the intermediate $U_0$ regime, the stripe ansatz evolves to the SQC$_1$ state, Fig.~\ref{fig2}(c). In Figs.~\ref{fig2}(d)-(f), we show the density patterns along the $y=0$ line where the enlargement of the TSS cluster triggering the transition to the SSS and SQC$_1$ states can be observed.
To some extent, stabilization of superfluid clusters and stripes is boosted by a process reminiscent of superfluids caged in puddles, as featured in BG phases~\cite{Fisher1989,PhysRevLett.103.140402,PhysRevLett.111.050406,soy11,PhysRevLett.131.173402,PhysRevB.105.094505}. 

In Fig.~\ref{fig3}, we show the behavior of quantities used to characterize the phases of the system and to locate their boundaries in Fig.~\ref{fig1}.
Additionally, we considered the density contrast to classify solid states~\cite{sm}.
We computed the superfluid fraction considering a generalization of Leggett's criterion~\cite{leggett1970},
\begin{align}\label{eq:global}
f =
\frac{L_x^2L_y^2}{\int d\mathbf{r_{\perp}} \rho_{\perp}(\mathbf{r_{\perp}})\int d\mathbf{r_{\perp}} \rho_{\perp}(\mathbf{r_{\perp}})^{-1}},
\end{align}
where $L_{x(y)}$ is the length of the system in the $x(y)$ direction, see Ref.~\cite{Coto2022}.
At strong $U_0$, the SQC transitions to the BG whenever $f = 0$.
To differentiate between a BG and a Mott insulator, we consider local superfluidity on the first ring of clusters 
around the origin~\cite{sm}. Both BG and insulating states have zero global superfluidity, but the 
former has local superfluidity in some puddles. The local superfluid fraction on a ring of radius $R$ is $f_{\text{loc}} = 4\pi^2/\int_0^{2\pi} Rd\theta\rho^{-1}_{\perp}(\mathbf{R})$, where $\mathbf{R} =(R\cos\theta,R\sin\theta)$ \cite{furutani2022}. 
In Fig.~\ref{fig3}(a), we show the local superfluid fraction and see that it remains finite when the global superfluidity vanishes ($U_0 \gtrsim 1.69$), which indicates that in the strong $U_0$ regime, the system is a BG.

At intermediate $U_0$, we observe a competition between the SSS (TSS) and QCL patterns. To measure how much each of these patterns contributes to the mixed state, we study the Fourier transform of the density $\tilde{\rho}_{\perp}(\mathbf{k}_{\perp}) = \int d\mathbf{r}_{\perp} \rho_{\perp} (\mathbf{r}_{\perp})e^{-i\mathbf{k}_{\perp}\cdot\mathbf{r}_{\perp}}$.
The SSS (TSS) and QCL patterns have, respectively, two (six) and eight characteristic peaks in the profile of $\tilde{\rho}_{\perp}(\mathbf{k}_{\perp})$. Once we measure the average height $h$ of each kind of characteristic peaks, see details in Ref.~\cite{sm}, we estimate the SSS (TSS) composition through the parameters $\Delta_{SSS(TSS)}=h_{SSS(TSS)}/(h_{SSS(TSS)} + h_{SQC})$ and $\Delta_{SQC}=h_{SQC}/(h_{SSS(TSS)} + h_{SQC})$~\cite{cha95}. The SSS (TSS) to SQC phase transition is thus identified as the lower bound of a phase in which $\Delta_{SQC}>\Delta_{SSS(TSS)}$, see Fig.~\ref{fig3}(b).

\begin{figure}[t!]
    \includegraphics[width=1.\linewidth]{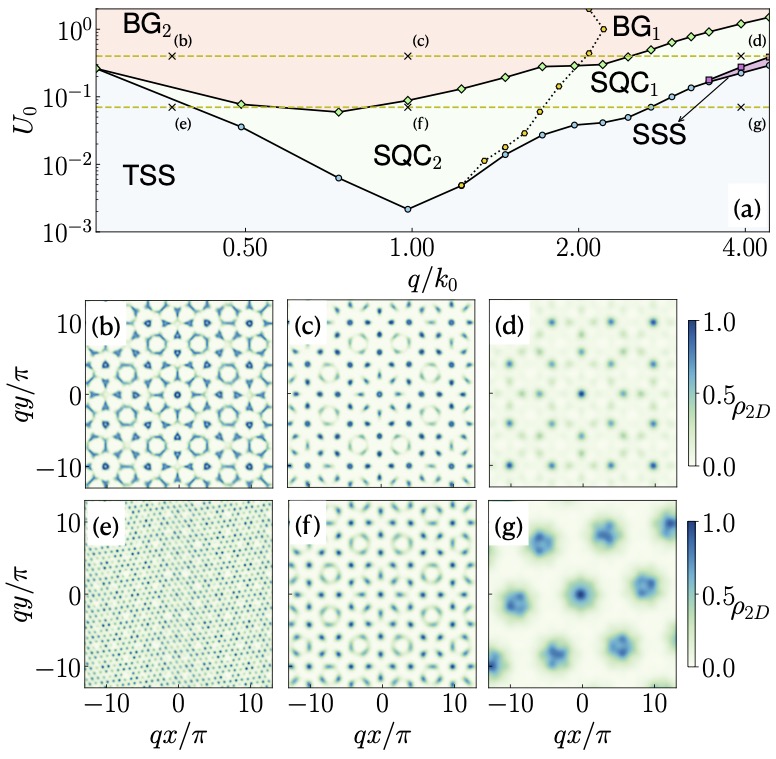}
    \caption{{\textbf{Phase diagram of dipolar bosons  in the $q$ versus $U_0$ plane.}}
    (a)  phase diagram obtained through the SVM at $\bar{\rho}_\perp=120$ and $a_s/a_{dd}=0.77$, we indicate with crosses the positions of the density patterns shown in figs.~(b)-(g).
    All the system parameters, but $q$, are the same as in Fig.~\ref{fig2}.
    (b)-(d) are three density patterns at $U_0 = 0.40$ and (b) $q = 0.15$, (c) $q = 0.4$, and (d) $q = 1.6$.
    (e)-(g) are three density patterns at $U_0 = 0.07$ and (e) $q = 0.15$, (f) $q = 0.4$, and (g) $q = 1.6$.
    }
    \label{fig4}
\end{figure}

The phenomenology discussed so far corresponds to a situation in which $q \gg k_0$.
The role of $q$ has been grasped by employing the SVM. Fig.~\ref{fig4}(a) depicts the phase diagram by varying $q/k_0$ and $U_0$, keeping both density $\bar{\rho}_{\perp}$ and $a_s/a_{dd}$ fixed.
For $q \lesssim k_0$, the TSS melts into the SQC$_2$ due to the competition with the QCL ordering for small values of $U_0$. In this case, the quasicrystalline pattern matches the QCL since the distance between neighboring sites is large enough to hinder the dipolar repulsion. Additionally, we also observe the BG$_2$ phase matching the lattice.
In panels (b)-(g), we show the ground-state configuration as $q$ increases for two different values of $U_0$. 
Importantly, we notice the existence of two opposite regimes of the TSS state. For $q\gg k_0$ we observe a triangular arrangement of clusters decorated by the QCL pattern (Fig.~\ref{fig4}(g)), while for $q\ll k_0$, the TSS state is structured as a quasicrystalline arrangement of clusters decorated by the triangular modulation (Fig.~\ref{fig4}(e)).
For strong $U_0$, the BG phase surprisingly melts in the superfluid SQC even if increasing $q$ is similar to increasing density. This phenomenology is also observed in contact-interaction systems~\cite{Gautier2021}.

\textit{Experimental feasibility} -- 
The results presented here are in a regime of parameters within current experimental capabilities~\cite{Chomaz2018,Kadau2016,Schmitt2016,Tanzi2019,Bottcher2019}. We propose a potential experiment using $^{162}$Dy atoms. This system offers a wide range of $s$-wave scattering lengths $a_s$, with a dipolar length $a_{dd}\approx 7\,\mathrm{nm}$, which results in a range of $a_s/a_{dd}$ consistent with the values considered in this work. 
For $^{162}$Dy, the characteristic units of length and time are $\ell=0.26\,\mathrm{\mu m}$ and 
$t_0=0.18\,\mathrm{ms}$. 
The trapping dimensionless frequency $\omega_z\, t_0=0.08$ is equivalent to  
$\omega_z\approx450\,\mathrm{Hz}$.
For $a_s/a_{dd}=0.77$ and effective two-dimensional density $\bar{\rho}_{\perp}=120$, we estimate a condensate thickness $\sigma_z \approx7.8\,\mathrm{\mu m}$ and a peak $3$D density in the absence of the QCL $3\rho_\perp/4\sigma_z\approx1.65\times10^{14}\,\mathrm{cm^{-3}}$. 
In these conditions, the ground-state characteristic wave vector of the triangular solid is $k_0\approx0.4$, resulting in a lattice spacing $\lambda=4\pi/(\sqrt{3}k_0)\approx 4.7\,\mathrm{\mu m}$, a small enough value to experimentally detect the long-distance properties of the predicted phases.

Recent works have successfully replicated the potential of an octagonal QCL using optical lattices~\cite{Viebahn2019, Sbroscia2020}. The experimental setup involves superimposing four coplanar 1D lattices, formed by retroreflective laser beams, arranged at $45^{\circ}$ angles relative to each other.
We considered QCLs whose characteristic wave vector $q$ approximately varies in the range from $k_0/4$ to $4k_0$, which results in a wavelength interval  $\Delta\lambda_q\approx(1.02,16.3)\mathrm{\mu m}$. For the lattice intensity $U_0$, our theoretical parameters also fall into an experimentally feasible region.
In~\cite{Sbroscia2020}, the authors use $U_0$ values up to $4.6 E_r$ ($E_r$ being the recoil energy), equivalent to $U_0 = 1.27$ in our units. This value falls within the BG phase in Fig.~\ref{fig4}.

We have considered finite-temperature corrections to our energy functional perturbatively~\cite{Baena_2023,Baena_2024}. Our calculations show that the phases observed remain robust for temperatures up to $T \sim 40$ nK (see SM). Regarding the effects of quantum fluctuations, it is known that the GPE description fails in the dilute regime~\cite{PhysRevResearch.1.033088}. Nonetheless, for the regime of parameters explored, corresponding to the high-density supersolid phase at $U_0=0$, we expect weak quantum correlations, therefore a valid GPE description. 
Moreover, the agreement between GPE calculations, Quantum Monte Carlo, and experiments has been attested in related scenarios~\cite{Biagioni2024,Cinti2014,Saito2016}.

\textit{Conclusions} -- 
In this work, we studied the effects of a QCL on the TSS phase exhibited by a planar system of dipolar bosons near the critical point. 
For weak QCL with high characteristic momentum, the system develops a triangular (TSS) or stripe (SSS) pattern decorated with a quasicrystalline structure that exhibits global superfluid properties. Interestingly, we observe that the TSS-to-SSS transition increasing the depth of the QCL occurs as a mechanism to minimize the dipolar repulsion when $q\gg k_0$.
A further increase of $U_0$ at the TSS or SSS phases eventually triggers a transition to the SQC phase and later to the BG phase. Moreover, by enlarging $q$, we observe that the quasicrystal phases develop a crossover from a state where the density pattern matches the QCL to a state in which they are different.
Finally, the insights from our investigation have laid the foundation for the experimental exploration, enabling us to understand the physics of long-range interacting quantum gases in quasi-periodic geometries.

\begin{acknowledgments}
{\it Acknowledgments:} We acknowledge M. Ciardi and Y.-C. Zhang for useful discussions. 
F. C. and V. Z. acknowledge financial support from PNRR MUR Project No. PE0000023-NQSTI. 
V. Z. thanks the Instituto Serrapilehira for its support (grant number Serra-1812- 27802).  A.M.C. acknowledges MPIPKS for financial support and hospitality.
We thank the High-Performance Computing Center (NPAD) at UFRN and  the NICIS Centre for High-Performance Computing, South Africa, for providing computational resources. 
\end{acknowledgments}

\bibliography{dipolar.bib}

\pagebreak
\widetext
\begin{center}
\textbf{\large Supplemental Materials: Exploring Supersolid Phases in Dipolar Quantum Gases through Quasicrystalline Confinement}
\end{center}

\setcounter{equation}{0}
\setcounter{figure}{0}
\setcounter{table}{0}
\setcounter{page}{1}
\makeatletter
\renewcommand{\theequation}{S\arabic{equation}}
\renewcommand{\thefigure}{S\arabic{figure}}

\section{I. Numerical solution of the GPE}

In the presence of dipolar interactions and for a weak QCL, several metastable states can be reached during the imaginary-time evolution. Therefore, the global minimum of the energy functional is found after comparing the energies of the stationary states resulting from imaginary-time evolution starting from different initial states. The numerical energy minimization was performed considering a system size consistent with the corresponding initial conditions and using periodic boundary conditions. The optimal solution for each kind of initial condition is found after minimizing the obtained energy concerning the seeded lattice constant $a$.     

We define $\psi(\mathbf{r},t=0) = \psi_z(z)\psi_{\perp}(\mathbf{r}_{\perp})$, with $\psi_z(z) \propto \exp[-z^2/(2\sigma_z^2)]$ and

\begin{align}\label{eq:ansatz}
\psi_{\perp}(\mathbf{r}_{\perp}) \propto \sum_i e^{-\frac{(x-x_i)^2+(y-y_i)^2}{2\sigma{\perp}^2}}.
\end{align}
The width of the functions in the sum of Eq.~(\ref{eq:ansatz}), initially defined by $\sigma_z$ and $\sigma_{\perp}$, and even their functional form, will evolve as the imaginary-time grows to render the smaller energy associated with a particular choice of the lattice sites $(x_i,y_i)$. We consider triangular and honeycomb lattices, known to be hexagonal solutions of the ground state of dipolar systems and perform computations with several lattice constants $a$ for each lattice. Additionally, we also consider a stripe ansatz $\psi_{\perp}(\mathbf{r}_{\perp}) \propto \exp[-\sum_ix_i^2/(2\sigma_{\perp}^2)]$, with $x_{i+1}-x_i = a$, and an homogeneous ansatz $\psi_{\perp}(\mathbf{r}_{\perp})\propto 1$.

\section{II. Effective 2D energy functional and variational ansatz}

The energy functional of the quasi-2D system is
\begin{align}\label{eqe}
    \dfrac{E[\psi]}{N} =\int d\mathbf{r}\Bigg\{{\dfrac{1}{2}|\nabla\psi(\textbf{r})|^{2}}+U(z)|\psi(\textbf{r})|^{2}+U_{q}(\mathbf{r})|\psi(\textbf{r})|^{2}\nonumber +\dfrac{2}{5}\gamma
    N^{3/2}|\psi(\textbf{r})|^{5} &+N\dfrac{a_{s}}{6a_{dd}}|\psi(\textbf{r})|^{4}\\
    &+\dfrac{N}{8\pi}\int d\mathbf{r'}V(\mathbf{r}-\mathbf{r'})|\psi(\mathbf{r})|^{2}|\psi(\mathbf{r'})|^{2}\Bigg\}.
\end{align}

With the Thomas-Fermi approximation for the $z$ component of the wave function, 
$\psi_0(z)=\sqrt{\frac{3}{4\sigma}\left(1-\frac{z^2}{\sigma^2}\right)}$, we can now integrate Eq.~\eqref{eqe} over the $z$-direction to obtain the 2D-projected energy-functional per particle, in terms of the in-plane wave function $\psi_\perp(\textbf{r}_\perp)$. Up to constant background contribution, this functional reads
\begin{align}
    \dfrac{\Delta E}{N} =\int \frac{d\mathbf{r}}{A}\Bigg\{{\dfrac{1}{2}|\nabla\psi_\perp(\textbf{r})|^{2}}+U_{q}(\mathbf{r})|\psi_\perp(\textbf{r})|^{2}\nonumber 
    +\dfrac{9\sqrt{3}\pi}{256\sigma^{3/2}}\gamma
   \rho^{3/2}|\psi_\perp(\textbf{r})|^{5} &+\dfrac{a_{s}}{10a_{dd}\sigma}\rho|\psi_\perp(\textbf{r})|^{4} \nonumber\\
    &+\dfrac{\rho}{2}\int d\mathbf{r'}V_{\mathrm{eff}}(\mathbf{r}-\mathbf{r'})|\psi_\perp(\mathbf{r})|^{2}|\psi_\perp(\mathbf{r'})|^{2}\Bigg\},
    \label{eq2}
\end{align}
where we omit the perpendicular symbol in all 2D position vectors to avoid a heavy notation. 
The effective interaction $V_{\mathrm{eff}}(\mathbf{r})$ produced by the dipolar interaction is such that it writes in momentum space as $\tilde{V}_{\mathrm{eff}}(k)=f(k\sigma)/\sigma$ where $f(x)=3/4\{[3-3x^2+2x^3-3(1+x)^2\exp(-2x)]/x^5 -4/15\}$.

The density configurations obtained from GPE (fig.~2) suggests a ground state wave function in which the periodic pattern envelopes the fast-varying quasiperiodic component of the wave function induced by the external lattice. In this way, we propose $\psi_\perp(\textbf{r})=\psi_{\mathrm{dip}}(\textbf{r})\psi_{\mathrm{oct}}(\textbf{r})$, where 
\begin{eqnarray}
\psi_{\mathrm{dip}}(\textbf{r})&=&\frac{1+\sum_{j\neq0}c_j\cos(\textbf{k}_j\cdot\textbf{r})/2}{\sqrt{1+\frac{1}{4}\sum_{j\neq0}c_j^2}},\\
\psi_{\mathrm{oct}}(\textbf{r})&=&\frac{1+\sum_{j\neq0}b_j\cos(\textbf{q}_j\cdot\textbf{r})/2}{\sqrt{1+\frac{1}{4}\sum_{j\neq0}b_j^2}}.
\end{eqnarray}
Here $\{c_j,\textbf{k}_j\}$ ($\{b_j,\textbf{q}_j\}$) represent the Fourier amplitudes and corresponding momenta of the periodic (quasiperiodic) component of the wave function. To ensure the expected symmetry properties of $\psi_{\mathrm{dip}}(\textbf{r})$ and $\psi_{\mathrm{oct}}(\textbf{r})$, Fourier amplitudes $c_j$'s or $b_j$'s
corresponding to wave vectors equivalent by symmetry operations of the respective component of the wave function are imposed to be equal. The resulting set of independent Fourier amplitudes as well as the lattice size $k_0$ of the momentum set $\{\textbf{k}_j\}$ are variationally determined. On the other hand, the "fractal" set of wave vectors $\{\textbf{q}_j\}$ is generated from the basis induced by the quasiperiodic octogonal external potential whose characteristic momentum is $q$.  In general, we should admit any possible orientation of the periodic component of the wave function concerning the external quasiperiodic lattice. To account for such degree of freedom we include an arbitrary orientation angle $\theta$ of the basis of the set $\{\textbf{k}_j\}$, measured with respect to the vector $\textbf{q}_1=q(1,0)$ imposed by the external potential. Naturally, the value of $\theta$ is determined variationally in our energy minimization process. To summarize the properties discussed so far, in Table 1 we present the basis wave vectors employed in the Fourier expansion of each component of the wave function.      

\renewcommand{\arraystretch}{2}
\begin{table}[h!]
\begin{tabular}{@{}|>{\centering\arraybackslash}p{.2
\textwidth}|>{\centering\arraybackslash}p{.3\textwidth}|c|@{}}
\cline{1-3}
    \textbf{Wave function component} & 
    \textbf{Basis vector} $\mathbf{k}_{0,j}$ & 
    \textbf{Index range} \\
\cline{1-3}
  $\psi_{\mathrm{dip}}$ (TSS)     &  
    $k_0(\cos(\frac{2\pi j}{6}+\theta),\sin(\frac{2\pi j}{6}+\theta))$ &  
    $j=0,1$ \\

    $\psi_{\mathrm{dip}}$ (SSS)   &  
    $k_0(\cos(\theta),\sin(\theta))$   &  
     \\

    $\psi_{\mathrm{oct}}$ &  
    $q(\cos\frac{2\pi j}{8},\sin\frac{2\pi j}{8})$ &
    $j=0,..., 3$\ \\ 
    \cline{1-3}
\end{tabular}
\caption{Components of the wave function for the different phases observed and their respective basis vectors}
\label{tpatterns}
\end{table}

For calculations, the set of Fourier modes $\{c_j,\textbf{k}_j\}$ is truncated at the fifth shell of the lattice, which is enough to observe convergence in the region of parameters investigated.
Additionally, the set of wave vectors $\{\textbf{q}_j\}$ is determined considering all combinations of up to five characteristic vectors of the quasiperiodic potential, which renders $52$ independent Fourier coefficients for the expansion of $\psi_\mathrm{oct}(\textbf{r})$. Finally, we verify upon minimization of the energy per particle for each kind of solution that $\theta=\pi/8$ corresponds to the optimal configuration for both the TSS and SSS patterns. However, the dependence of the energy per particle with $\theta$ is extremely weak in the regime of parameters explored.

\begin{figure}[h]
    \centering
    \includegraphics[width=0.5\linewidth]{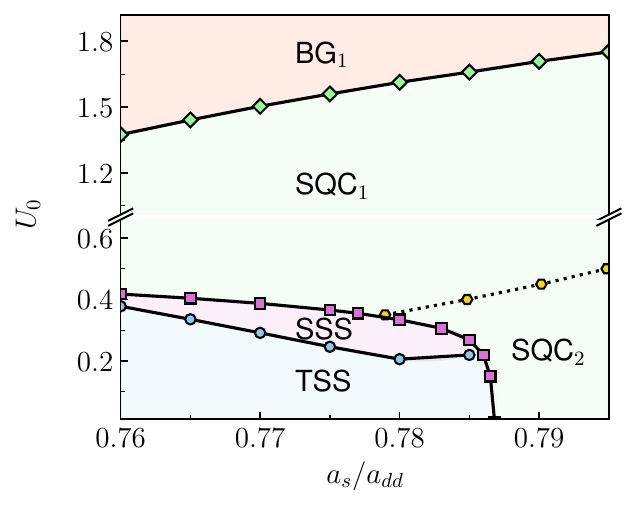}
    \caption{Phase diagram for dipolar bosons in an external quasiperiodic lattice with $q=1.8$, varying $a_s/a_{dd}$ and $U_0$ at a fixed average planar density $\bar{\rho}_\perp=120$. The diagram was calculated through the spectral variational method described above. As in Fig.~1 we detect the TSS, SSS, SQC$_n$ and BG$_n$ phases. Where, for $n = 1$ the QC pattern is different from the QCL and for $n = 2$ they are equal.}
    \label{fig_supmat_3}
\end{figure}

The obtained $U_0$ versus $a_s/a_{dd}$ phase diagram using the previously discussed spectral variational method is shown in Fig.(\ref{fig_supmat_3}). We considered a QCL with $q=1.8$ and an average 2D density of particles $\bar{\rho}_\perp=120$. The obtained phase diagram coincides qualitatively with the one shown in Fig.~1. The small deviations observed are expected to be produced by the differences in the working model, i.e. the analytical work was done using a 2D effective model. Additionally, the variational approach implemented is also limited by the particular ansatz considered for the different phases as well as the GPE is also affected by finite size effects.

\section{III. Energy cost of the relative orientation of SSS and QCL}
\begin{figure}[h!]
    \centering
    \includegraphics[width=0.5\linewidth]{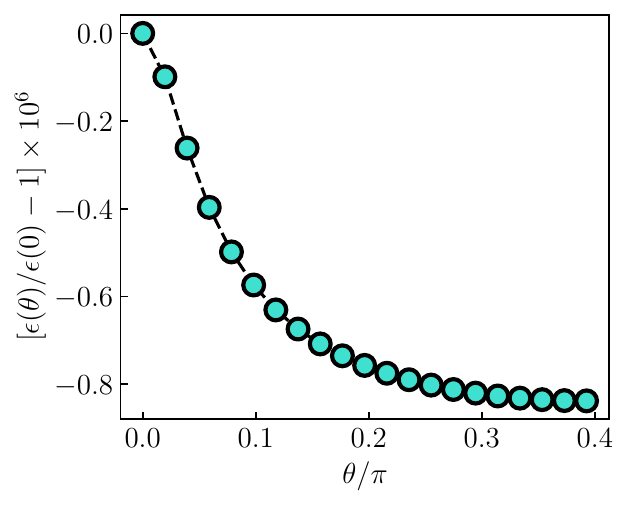}
    \caption{Angular dependence of the energy per particle of the SSS phase varying the orientation of the stripes pattern with respect to one of the main directions of the QC lattice. Results were obtained through the variational spectral method considering the parameters $U_0=0.25$, $q=1.8$ $a_s/a_{dd}=0.78$ and $\rho_\perp=120$. }
    \label{figs5}
\end{figure}

As mentioned previously the variational spectral method implemented considers an arbitrary orientation of the TSS and SSS with respect to the QCL. Such orientation is parameterized via the angle $\theta$ previously defined. Our variational study lead us to the conclusion that the optimal orientation of both phases is reached for  $\theta=\pi/8$. We note that such a configuration would be difficult to implement in the GPE solution due to the periodic boundary conditions. To study the impact of setting $\theta=0$, as is the GPE, we calculate the relative energy per particle difference $[\epsilon(\theta)-\epsilon(0)]/\epsilon(0)$  for a typical configuration within the SSS phase. The results are presented in Fig~(\ref{figs5}), as we can observe the relative error in such approximation is of order $10^{-6}$. Such a small impact on the energy cost of the SSS and TSS implies that in this regard the GPE in the regime of parameters considered are unaffected by the approximation of setting $\theta=0$.

\section{IV. SQC pattern different from the external lattice}

\begin{figure}[h]
    \centering
    \includegraphics[scale=0.8]{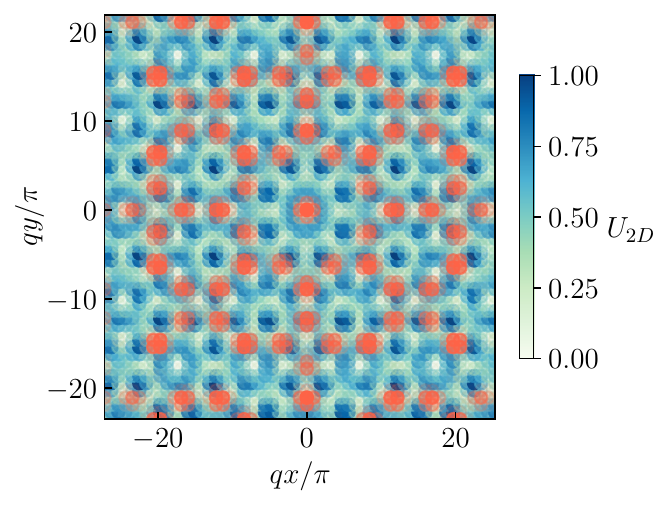}
    \caption{{\bf SQC versus QCL}. The red dots represent the normalized density pattern of the SQC shown in Fig.~2(c).
    Notice that not all the minima of the QCL $U_{2D}=(U_q - U_q^{\min})/(U_q^{\max}-U_q^{\min})$ are populated.}
    \label{figx}
\end{figure}

In Fig.~\ref{figx}, we show the QCL and the density pattern of the SQC in Fig.~2(c). We see that not all the minima of the QCL are populated. This is a mechanism to minimize the dipolar repulsion in the ground state when $q \gg k_0$.

\section{V. Contrast}

\begin{figure}[h]
    \centering
    \includegraphics[width=0.5\linewidth]{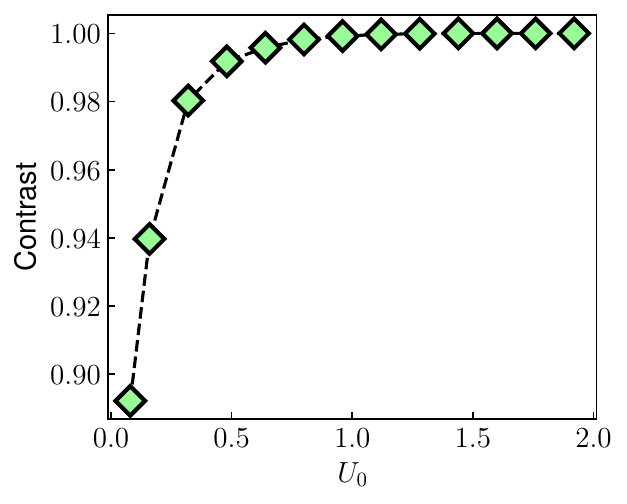}
    \caption{Density contrast as a function of $U_0$ for $\bar{\rho}_{\perp}=120$, $a_s/a_{dd}=0.78$, $\omega_z = 0.08$ and $q=1.8$.}
    \label{fig:c}
\end{figure}

The contrast is defined as $\mathcal{C}=(n_{\text{max}}-n_{\text{min}})/(n_{\text{max}}+n_{\text{min}})$,
with $n_{\text{max(min)}}$ being the maximum (minimum) value of $\rho_{\perp}(\mathbf{r}_{\perp})$.
In Fig.~\ref{fig:c}, we obtained finite values of $\mathcal{C}$, ruling out the existence of a homogeneous liquid phase in the regimes we considered in Fig.~3 of the main text.

\section{VI. Locating rings with a finite density of particles}

\begin{figure}[h]
    \centering
    \includegraphics[scale=0.8]{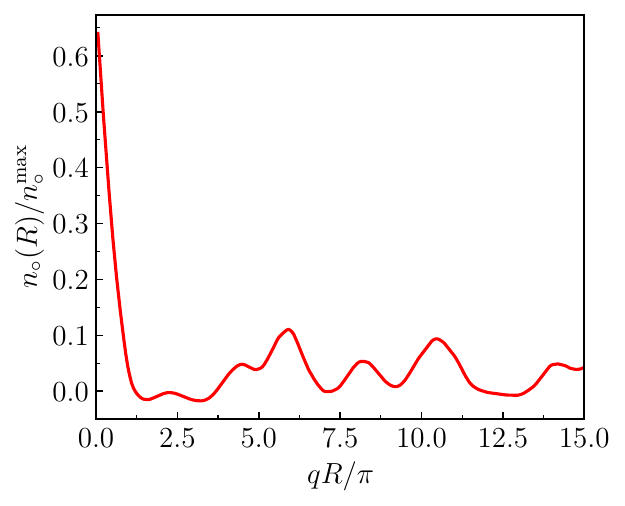}
    \caption{The average number of dipolar bosons in rings of radius $R$. The average on-plane density is $\bar{\rho}_{\perp}=120$, the ratio $a_s/a_{dd}=0.78$ and $U_0 = 0.56$.}
    \label{fig:nring}
\end{figure}

To locate the rings where local superfluidity is considered, we compute $n_{\circ}(R) = \int d\theta \rho_{\perp}(\mathbf{R})/2\pi$, with $\mathbf{R}=R(\cos\theta,\sin\theta)$. Results for this quantity are shown in Fig.~\ref{fig:nring}, where we considered $a_s/a_{dd}=0.78$ and $U_0=0.56$. We see that the first ring around the origin is located at $R \sim 6$, a value that does not change in the regime of $a_s/a_{dd}$ we have considered.

\section{VII. Locating and measuring the peaks of $\tilde{\rho}_{\perp}(\mathbf{k}_{\perp})$}

\begin{figure}[h]
    \centering
    \includegraphics[width=\linewidth]{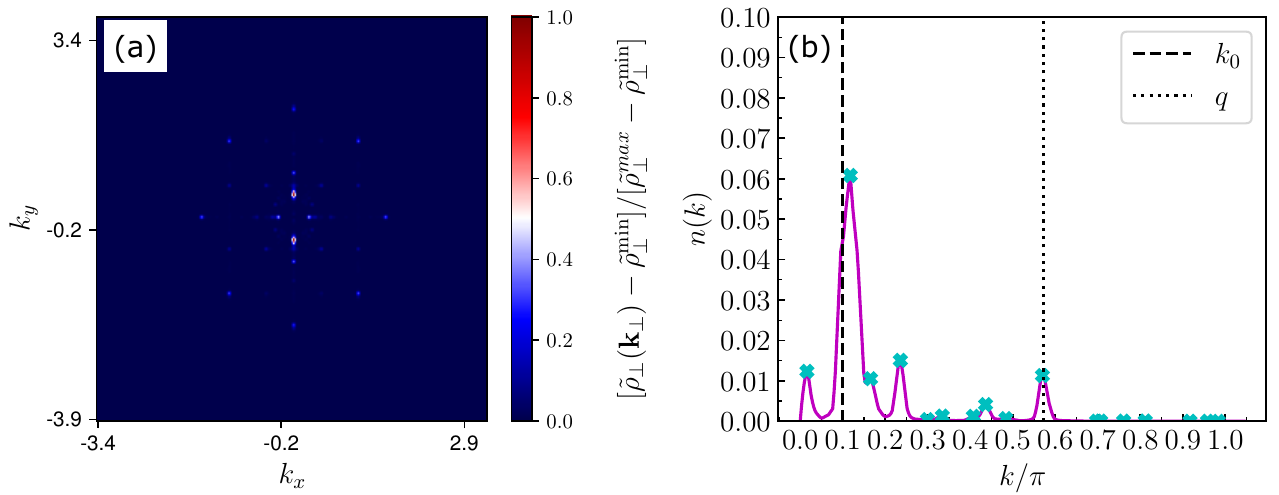}
    \caption{(a) Density pattern in reciprocal space and (b) angle averaged density in the reciprocal space. In $k = k_0$, we have the characteristic momentum of the stripes due to the dipolar interaction($k_0= 2\pi/a \approx 0.3$). In $k = q = 1.8$, we have the characteristic momentum of the underlying quasiperiodic lattice. The peak at $k=0$ was removed. The system parameters are $\omega_z = 0.08$, $\bar{\rho}_{\perp}=120$, the ratio $a_s/a_{dd}=0.78$ and $U_0 = 0.42$.
    Peaks were found at $k$ = $0.05$, $0.31$, $0.53$, $0.74$, $1.04$, $1.28$, $1.37$, $1.79$, $2.40$,  $2.55$, $2.88$, and $3.09$.}
    \label{fig:densk}
\end{figure}

In Fig.~\ref{fig:densk}(a), we show the density profile in reciprocal space for situations where the quasiperiodic pattern dominates. The octagons shown have different sizes and orientations, therefore we devise a procedure to locate and measure the height of the most intense peaks.

We compute the angle-averaged density in momentum space $n_{\circ}(k)= \int d\theta \tilde\rho_{\perp}(\mathbf{k})/2\pi$, with $\mathbf{k}=k(\cos\theta,\sin\theta)$, which is shown in Fig.~\ref{fig:densk}(b).
We see that peaks appear in the neighborhoods of $k = k_0$ and $k = q$, which are related to the characteristic modulation of the SSS (TSS) and SQC patterns respectively.  Then, we locate the peaks $k = k_{\text{peaks}}$ of the circumference circumscribing structures with two (SSS), six (TSS), or eight (SQC) peaks, and we search for the structure that fits our data the best. To do so, we compute

\begin{align}\label{eq:sum_peaks}
    S_{m}(k_{\text{peaks}},\theta_0) = \sum_{\mathbf{k}_{\text{peaks}}}\tilde{\rho}_{\perp}(\mathbf{k}_{\text{peaks}}),
\end{align}
with $\mathbf{k}_{\text{peaks}}=k_{\text{peaks}}(\cos\theta,\sin\theta)$, $\theta = \theta_0 + n\pi/m$, $n$ being an integer in the interval $[0,m[$, and $m=2,6$, and $8$ respectively for the SSS, TSS and SQC states. We span over several values of $\theta_0$ to consider many possible orientations of each structure. The best fit to our data provides the maximum value $S_{m}^{\max}$, therefore the height of the peaks can be computed as $h=S_{m}^{\max}/m$.

\section{VIII. Fourier components of the QC phase}

In this section, we present some details of a typical QC configuration obtained by using the SVM. Although we have considered a large Fourier basis for the QC pattern we observed that only a few Fourier modes are activated by the system. To illustrate this fact, we present the results for the five more relevant Fourier modes of a typical configuration within the QC phase for $q=1.6$ and $\rho_\perp=120$. It can be verified that as observed in Fig.\ref{fig:densk} the wave vectors corresponding to the three most relevant modes are approximately located at $q$, $0.765 q$, and $0.414 q$. 

\renewcommand{\arraystretch}{2}
\begin{table}[h!]
\begin{tabular}{@{}|>{\centering\arraybackslash}p{.26
\textwidth}|>{\centering\arraybackslash}p{.12\textwidth}|>{\centering\arraybackslash}p{.12\textwidth}|>{\centering\arraybackslash}p{.12\textwidth}|>{\centering\arraybackslash}p{.12\textwidth}|>{\centering\arraybackslash}p{.12\textwidth}|>{\centering\arraybackslash}p{.12\textwidth}@{}}
\cline{1-6}
\multicolumn{6}{|l|}{\textbf{Structure Factor QC, $U_0=1.2$, $a_s/a_{dd}=0.77$, $\rho_\perp=120$, $q=1.6$}}\\ \hline
\cline{1-6} 

\textbf{Momentum Modulus ($\vert q\vert$)} & $1.000$ & $0.765$ & $0.414$ & $1.414$ & $1.082$ \\ \hline
\cline{1-6} 

\textbf{Momentum Orientation ($\theta$)} & $0$ & $\pi/8$ & $0$ & $0$ & $\pi/8$ \\ \hline

\cline{1-6}
\textbf{Amplitudes $(\tilde{\rho}(\vec{q}))$} & $0.382$ & $0.244$ & $0.174$ & $0.077$ & $0.035$ \\ \hline
\end{tabular}
\caption{Main Fourier components of the normalized QC density pattern. Momenta are expressed in units of the characteristic momentum of the external lattice $q$. The momentum orientation is relative to the $x$ axis.}
\end{table}
 \newpage
\section{IIX. Free energy corrections at finite temperatures.}
At finite temperatures, the free energy functional of the system must be corrected due to the presence of thermal fluctuations. Perturbation theory allows us to obtain systematically such corrections if we depart from the coherent-state path-integral partition function for a system of interacting bosons. For a comprehensive deduction of the leading quantum and thermal corrections to the free energy, see section 11.4 of ref.~\cite{Stoof}, where the case of a boson gas with contact interaction is studied in detail. The generalization of this result to the case of a homogeneous polarized dipolar bosonic gas is straightforward leading us to the expression~\cite{Baena_2023,Baena_2024}
\begin{equation}
\Delta F=V\int \frac{d^3k}{(2\pi)^3}\frac{1}{\beta}\ln\left(1-e^{-\beta\epsilon(\textbf{k},\rho)}\right),
\end{equation}
where $\epsilon(\textbf{k},\rho)=\sqrt{\frac{k^2}{2}\left[\frac{k^2}{2}+2\rho\left(\frac{1}{3}(a_s/a_{dd}-1)+\cos^2\theta\right)\right]}$ and $\cos\theta=k_z/k$. As a consequence, within the Local Density Approximation, the finite temperature correction to the total free energy per particle will be 
\begin{equation}
\Delta F/N=\Delta f=\int \frac{d^2r}{A}\int_{-\sigma}^{\sigma} dz\int \frac{d^3k}{(2\pi)^3}\frac{1}{\rho_\perp\beta}\ln(1-e^{-\beta\epsilon(\textbf{k},\rho(\textbf{r}))}).
\end{equation}
In the cases in which $a_s/a_{dd}<1$, $\epsilon(\textbf{k},\rho)$ becomes complex at low momenta, signaling the instability of the homogeneous state. To deal with this issue, a low momentum cut-off scheme have to be selected to limit the characteristic wave length fluctuations to a region in which $\epsilon(\textbf{k},\rho)$ is positive~\cite{Bisset2016}. For consistency with the expression employed for the LHY correction, we consider the integration over all momenta satisfying $\epsilon(\textbf{k},\rho)>0$. After a lengthy but direct calculation, it is possible to show that 
\begin{equation}
\Delta f[\rho_\perp(\textbf{r}_\perp)]=\frac{1}{\beta\rho_\perp}\int\frac{d^2r}{A}\left[-\frac{\sigma^{3/2}\zeta(3)}{2\sqrt{3}\beta^2\sqrt{\rho_\perp(\textbf{r}_\perp)}}+\frac{(2+a_s/a_{dd})^2\rho_\perp(\textbf{r}_\perp)^{3/2}}{16\sqrt{3}\pi^2\sigma^{1/2}}G\left(\frac{a_s}{a_{dd}},\frac{\beta(2+a_s/a_{dd})\rho_\perp(\textbf{r}_\perp)}{4\sigma}\right)\right],
\label{fintemp}
\end{equation}
where
\begin{equation}
G(a,b)=2\int_0^1dz\int_0^\infty dy \frac{y}{\sqrt{1-z^2}}\arctan\left[\frac{3(1-z^2)-(2+a)\sqrt{(1-z^2)^2+y^2}}{\sqrt{6(2+a)\left(\sqrt{(1-z^2)^4+(1-z^2)^2y^2}-(1-z^2)^2\right)}}\right]\ln\left(1-e^{-by}\right)
\end{equation}
and $\zeta(x)$ represent the Riemann zeta function. 

Now, the finite temperature free energy functional can be obtained by adding the above correction to the ground state energy functional in Eq.~\eqref{eqe}. In principle, to study the phase diagram of our system for $T>0$ we should proceed with the minimization of the free energy functional obtained for the different phases detected in the ground state phase diagram. However, for a simple estimation of the phase diagram in the low-temperature region, we can follow an alternative perturbative approach. The finite temperature free energy of the different phases can be estimated adding to the corresponding ground state energy the value of the correction $\Delta f[\rho_{\perp}(\textbf{r}_\perp)]$ evaluated at the corresponding zero temperature optimized solution. In this way, the values of the free energy of each solution is obtained for $T>0$ allowing us to calculate the phase diagram.   

Within this approximation, we reached to the following finite temperature phase diagram in the $U_0$ versus $k_BT/\epsilon$ plane at $\omega_z=0.08$, $\bar{\rho}_\perp=120$, $a_s/a_{dd}=0.78$ and $q=1.8$. For the case of dysprosium atoms considered in the main text, the value of the energy scale $\epsilon$ can be directly determined, and consequently it is possible to map the dimensionless temperature axis to real temperatures for the proposed experiment. As can be observed, the phases obtained at zero temperature are stable against thermal fluctuations and they remain to exist at temperatures well within our current experimental capabilities.    

\begin{figure}[h]
    \centering
    \includegraphics[width=0.5\linewidth]{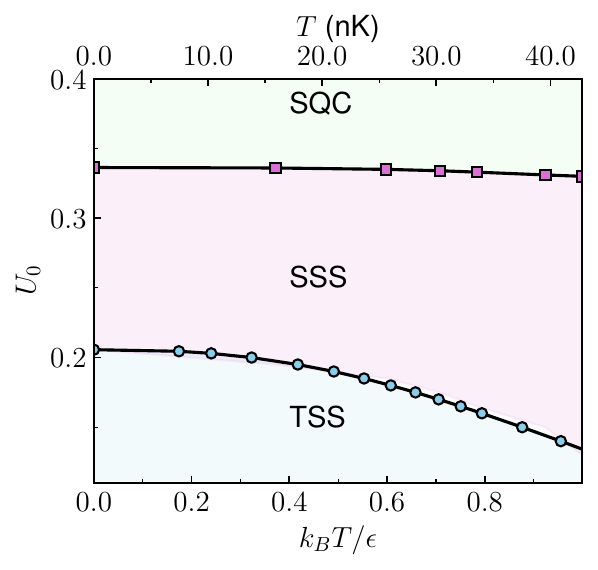}
    \caption{$U_0/\epsilon$ versus $k_BT/\epsilon$ phase diagram for the potential experiment considered in the main text.}
\end{figure}

\section{IX. Amplified density patterns.}

\begin{figure}[h]
    \centering
    \includegraphics[width=1.0\linewidth]{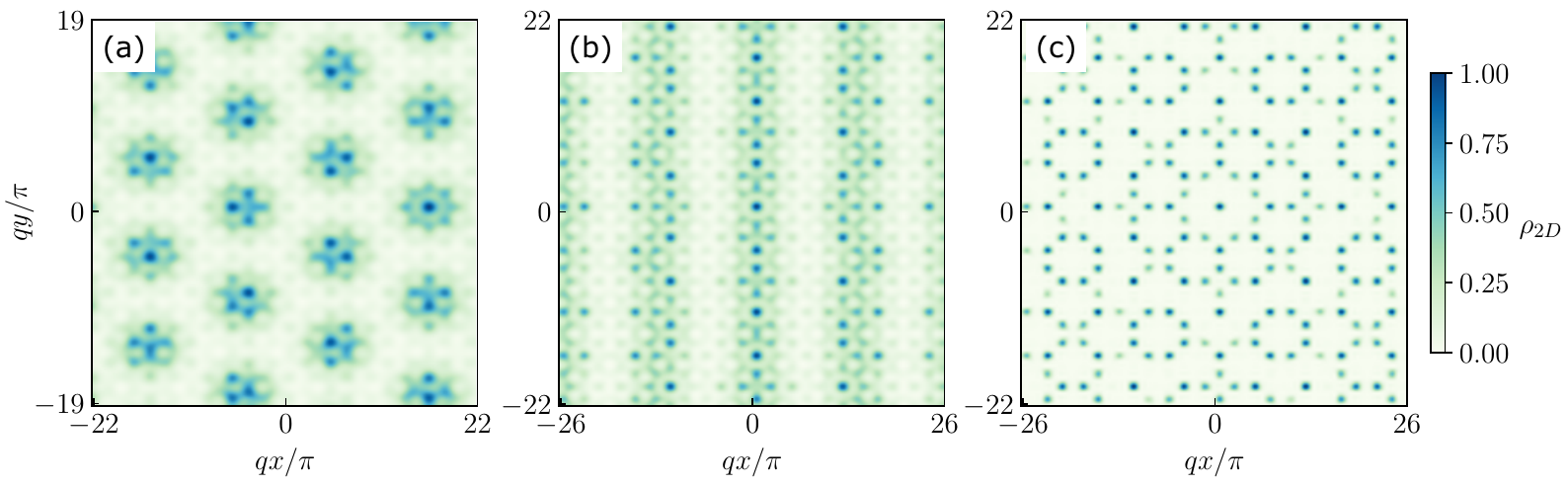}
    \caption{Amplification of density patterns in Fig.~2(a)-(c)}
    \label{s8}
\end{figure}

The clusters of the TSS, SSS, and SQC state are shown enlarged in Fig.~\ref{s8}.

\end{document}